\documentclass{PoS}

\title{Constraints on neutron-star theories from nearby neutron star observations}

\ShortTitle{Observational constraints for EoS}

\author{\speaker{Ralph Neuh\"auser}\thanks{We thank the german national science foundation (DFG) in 
SFB TR7 on graviational wave astronomy for support. NT also thanks the Carl-Zeiss Stiftung for support.}\\
        Astrophysical Institute, University Jena, Schillerg\"asschen 2, D-07745 Jena, Germany\\
        E-mail: \email{rne@astro.uni-jena.de}}

\author{Valeri V. Hambaryan \\
        Astrophysical Institute, University Jena, Schillerg\"asschen 2, D-07745 Jena, Germany\\
        E-mail: \email{vvh@astro.uni-jena.de}}

\author{Nina Tetzlaff\\
        Astrophysical Institute, University Jena, Schillerg\"asschen 2, D-07745 Jena, Germany\\
        E-mail: \email{nina@astro.uni-jena.de}}
        
\author{Markus M. Hohle\\
        Astrophysical Institute, University Jena, Schillerg\"asschen 2, D-07745 Jena, Germany\\
        MPI f\"ur extraterrestrische Physik, Giessenbachstrasse 1, D-85741 Garching, Germany \\
        E-mail: \email{mhohle@astro.uni-jena.de}}
        
\author{Thomas Eisenbeiss\\
        Astrophysical Institute, University Jena, Schillerg\"asschen 2, D-07745 Jena, Germany\\
        E-mail: \email{eisen@astro.uni-jena.de}}

\abstract{We try to constrain the nuclear Equation-of-State (EoS)
and supernova ejecta models by observations of young
neutron stars in our galactic neighbourhood.
There are seven thermally emitting isolated neutron stars known from X-ray
and optical observations, the so-called Magnificent Seven,
which are young (few Myrs), nearby (few hundred pc), and
radio-quiet with blackbody-like X-ray spectra, so that 
- by observing their surface - 
we can determine their luminosity, distance, and temperature,
hence, their radius. We also see the possibility to determine
their current neutron star masses and the masses of their progenitor stars
by studying their origin.
It is even feasible to find the neutron star which was born
in the supernova, from which those Fe60 atoms were ejected,
which were recently found in the Earth crust.}

\FullConference{11th Symposium on Nuclei in the Cosmos, NIC XI\\
		July 19-23, 2010\\
		Heidelberg, Germany}

\begin{document}

\section{Introduction: The Magnificent Seven Neutron Stars}

A small group of very intriguing Neutron Stars (NS) were discovered by ROSAT, 
they are sometimes called ``X-ray (bright, optical) Dim Isolated Neutron Stars''
(XDINS), given that they are X-ray bright and optical dim,
or the ``Magnificent Seven'' (M7), because there are seven
such objects known. All seven show thermal X-ray
spectra characterized by blackbody temperatures 
$T^\infty_{\rm bb}\approx 0.5-1.1$ MK (Table~1).
Given their X-ray periods and period derivatives (Table~1),
their characteristic ages are around a few million years.
The M7 were probably born in the Gould Belt 
(see Poppel 1997 for a review),
a torus-like structure of star forming regions 
and thousands of low- to high-mass stars up to 50 Myr young
(e.g. Neuh\"auser 1997),
all within 0.6 kpc. The NS birth rate may be larger in the Gould Belt
(Grenier 2000, Popov et al. 2005, Posselt et al. 2008) 
than in the surrounding field (Tammann et al. 1994),
namely about 30 SN/Myr in the Gould Belt
(Hohle et al. 2010a, Grenier et al. 2000).

\begin{table}
\small
\label{tbl}
\begin{tabular}{lccccccc}
\multicolumn{8}{c}{Table 1: Isolated thermaly emitting Neutron Stars (Magnificent Seven)} \\ \hline
Object  & Temperature     & Rotation & $dP/dt$  &  Pulse   & Distance          & Optical        & Proper       \\
        & $kT_{\infty}$   & period   &($10^{-14}$)& fraction &                    & magnitude      & motion       \\
        & [eV]            & [sec]    &[sec/sec]   & [\%]     & [pc]               & [mag]          & [mas/yr]     \\ \hline
RXJ1856 & $63$            & 7.05     &      3     & 1        & $123 \pm 13$       & V=25.7         & $332 $       \\
RXJ0720 & $85-95$         & 8.39     &      7     & 8-11     & $360^{+170}_{-90}$ & V=26.8         & $108$        \\
RXJ1605 & $96$            & 6.88     &     ---    & $\le 3$  & $\le 410$          & B=27.2         & $155$        \\
RXJ0806 & $96$            & 11.37    &      6     &  6       & $240 ^{+10} _{-5}$ & B$\ge 24$      & $\le 86$  \\
RXJ1308 & $102$           & 10.31    &     11     & 18       & 76-380& m$_{\rm 50CCD}=28.6$  & $223 $       \\
RXJ2143 & $100$           & 9.44     &      4     &  4       & $\ge 250$          & B=27.9         & ---          \\
RXJ0420 & $44$            & 3.45     &     ---    & 17       & $\approx 345$      & B=26.6         & $\le 123$  \\ \hline  
\end{tabular} \\
Remarks: RXJ0720 has variable temperature and pulse period (possibly due to precession or glitch); 
only the distances of RXJ1856 and RXJ0720 are measured directly as trigonometric parallaxe, 
while the other distances are estimated from X-ray flux and absorbing column density; 
proper motion in milli arc seconds per year [mas/yr]. References: 
Hohle et al. 2009, Haberl 2007, Burwitz et al. 2003, van Kerkwijk \& Kaplan 2008, 
Kaplan et al. 2007, Eisenbeiss et al. 2010, Schwope et al. 2007, 2009, 
Motch et al. 2008, 2009, Haberl et al. 2004, Walter et al. 2010.
\end{table}

For RXJ1856.5-3754 (or RXJ1856 for short), the first and brightest M7 
(Walter, Wolk, and Neuh\"auser 1996), a trigonometric parallax was
obtained by the Hubble Space Telescope (HST) yielding $117 \pm 12$ pc
(Walter \& Lattimer 2002). While Kaplan and van Kerkwijk argue for a larger
distance of up to $\approx 161-167$~pc (Kaplan et al. 2007, van Kerkwijk \& Kaplan 2007), 
we have re-evaluated available HST data and found $123 \pm 13$ pc (Walter et al. 2010).  
The X-ray spectral energy distribution gives a surface temperature
$kT = 60$ eV. With the Stefan-Boltzmann law and under consideration of both the
X-ray spectrum and the optical spectrum (both blackbodies), this results in a
$R_{\infty}= 17$ km (Pons et al. 2002, Tr\"umper et al. 2004, Walter et al. 2010).
This is consistent with a stiff EoS at high densities, 
excluding quark stars, which was suggested before based on a
preliminary (but incorrect) smaller distance (Drake et al. 2002). This already
constitutes a constraint for the EoS. 
Both optical and X-ray spectra of this NS were recently described 
by a thin, partially ionized hydrogen atmosphere above a
condensed surface and with high magnetic field ($B \approx 3$-$4 \times 10^{12}$
G, Ho et al. 2007). At 140 pc, this also gives $R_{\infty} = 17$ km.
RXJ1856 is an isolated NS in the sense that there is neither a supernova (SN) remnant nearby
nor a companion detected. The upper mass limit for undetected companions is 16 Jupiter
masses from deep infrared imaging (Posselt et al. 2009). 

For RXJ0720, the HST parallax gives $\approx 360$ pc
(Kaplan et al. 2007), so its radius can be determined, too (Eisenbeiss 2010).
It shows a long-term X-ray variability on a time-scale of 7-14 yrs,
which might be either precession or a glitch
(Haberl et al. 2006, van Kerkwijk et al. 2007, Hohle et al. 2009, 2010b).

We know the distances of RXJ1856 and RXJ0720 and directly observe their surface
emission, so that we can determine their true radii $R$ (model-dependent). What is
missing for a stringent constraint on the EoS is an estimate of the mass $M$. Depending on
the atmospheric composition and the magnetic field strength, one would expect
certain lines in spectra (cyclotron lines or atomic absorption lines) observable
with XMM-Newton and Chandra. Both gravitational redshift and
surface gravity depend on $M$ and $R$, so that a measurement of either of the
two or both would yield $M$ (together with the known radius). Very surprisingly,
no absorption lines were found in the spectra of RXJ1856
(Burwitz et al. 2001, 2003), so that neither redshift nor
gravity could be determined, yet. It is still a matter of debate, as to why
there may be no (apparent) absorption lines in RXJ1856, e.g. vacuum-polarization
effects, very high magnetic fields, or a condensed surface
(Tr\"umper et al. 2004, Ho \& Lai 2007, Ho et al. 2007). 

Broad absorption features at 0.3-0.8 keV were detected in six objects, 
whereas RXJ1856 reveals almost a ``perfect'' blackbody spectrum. 
The origin of the absorption features might be caused by
proton-cyclotron resonance in a magnetic field ($B\approx 3\times 10^{13}$~G) or
produced by atomic transitions in a strongly magnetized hydrogen surface.
But the features could also stem from ion-cyclotron absorption,
or atomic transitions in neutral hydrogen or heavy-metal ions.
For more information on the M7, see recent reviews in
Haberl (2007) and van Kerkwijk \& Kaplan (2007).

Observational studies of thermal spectra and comparing it with theoretical
models of highly magnetized NS atmospheres is a promising way to improve our
understanding of emission processes and to achieve constraints on the EoS.

\section{Constraining the Equation-of-State}

{\bf Radius from luminosity and temperature.}
With several HST images of the brightest M7, RXJ 1856, 
one could determine its proper motion (1/3 arc sec/year,
Walter 2001) and its parallax, i.e. distance ($123 \pm 13$ pc), 
i.e. with a precision of about $10~\%$ (Walter et al. 2010).
Distance and brightness (also known from the HST imaging) 
together yield the luminosity (to within $\approx 20~\%$).

The Chandra and XMM X-ray spectra of RXJ1856 does not show any deviation from a blackbody
(Burwitz et al. 2001, 2003, Pons et al. 2002).
The spectrum yields the temperature of the thermally emitting hot surface
(energy kT = 63 eV), to within $5~\%$ (Burwitz et al. 2001, 2003). 
Luminosity and temperature together give the radius of the emitting area (to within $\approx 25~\%$).
For all M7, the continuum of the X-ray spectrum is not
consistent with the optical data; the latter show an excess 
by a factor of about 7 for RXJ1856 (Burwitz et al. 2003).
This indicates that the hot X-ray emitting area (polar spots ?) is smaller 
than the remaining (cooler, but still warm) surface of the NS, which emits in the optical.
The best fitting model for both the X-ray and optical data can then yield
the radius to be $R_{\infty} = 17$ km (Tr\"umper 2003, Walter et al. 2010). 
This radius is therefore model-dependent.

The mass of such a neutron star can in principle be determined from Kepler´s 3rd law, 
if it is orbited by a low-mass companion (but no such companions were found 
down to brown dwarf masses, Posselt et al. 2009). 
Another possibility would be a gravitational microlensing event, when the NS
happens to move exactly across a distant background star (Paczynski 2001); however,
no such stars are detected on the flight path of the M7 for the whole 21st 
century (Eisenbeiss 2010).
 
{\bf Compactness from phase-resolved spectroscopy.}
Almost all isolated NSs have an absorption feature in their X-ray spectrum. 
By considering various theoretical models for the magnetized surface like
naked condensed iron surfaces and partially ionized hydrogen model atmospheres
(Suleimanov et al. 2009), the absorption features and the angular
distributions of the emergent radiation can be studied. 

For computing light curves and integral emergent spectra of magnetized NSs, 
we use a new neutron star model atmosphere with general relativistic 
corrections (Suleimanov et al. 2010). We are currently testing a new model 
for phase-resolved spectroscopy for the 
case of the M7 with the largest pulse fraction, i.e. RXJ1308 = RBS 1223. We
assumed a dipole surface magnetic field distribution with a possible toroidal
component and corresponding temperature distribution. The model with two uniform
hot spots at the magnetic poles (or with a small shift angle, i.e. not
exactly opposite of each other) is employed. Light curves and spectra
were computed using different surface temperature distributions and various
local surface models. We concluded that in order to explain the prominent
absorption features in the soft X-ray spectra of isolated NSs (especially in
RBS 1223), a thin atmosphere above the condensed surface must be invoked. 
A strong toroidal magnetic field component on the stellar surface can be excluded.
If the toroidal component of the magnetic field on the surface is 3 to 7
times larger than the poloidal component, the absorption feature in the integral
spectrum is too wide and shallow to be detectable.
By a simultaneous fit (Markov Chain Monte Carlo method) to both the light curves
at 200 different energies and to the spectra at 20 different rotational 
phases, i.e. by doing rotational phase-resolved spectroscopy, 
we can fit and, hence, determine several parameters of the NS 
emission, e.g. sizes and temperatures of two hot
polar spots, the geometry of the NS, the magnetic field strength, the strength 
of the cyclotron emission, and also the gravitational redshift. 
The latter depends on the compactness
of the NS, i.e. gives a constraint on the ratio of mass over radius.

This new method is currently being applied to RBS1223,
which has the largest pulse fraction and a double-hamped light curve, i.e. two
different hot polar spots (Hambaryan et al., in preparation). The value on the
compactness can be obtained with a precision of $10-20~\%$. 
 
Phase-resolved spectroscopy will then
be applied to the other M7.
In the case of RXJ1856, where the radius has been determined previously by a
different technique, which we currently also apply to RXJ0720, a newly determined
compactness and the known radius together give the mass, for the first time the mass
of an isolated neutron star, where no mass exchange could have happened.
With a radius to within $\approx 25~\%$ and a compactness to within $\approx 15~\%$,
we can determine the mass to better than within a factor of 2.

\section{Identifying birth places of neutron stars}

NSs obtain very large space motions (kicks)
during their birth in a SN event 
(Burrows \& Hayes 1996, Janka \& Mueller 1996, Wang et al. 2006),
so that they have large space velocities
(Lyne \& Lorimer 1994, Lorimer et al. 1997, Hansen \& Phinney 1997, Cordes \& Chernoff 1998,
Arzoumanian et al. 2002, and Hobbs et al. 2005), 
and therefore large proper motions if nearby.
About $27~\%$ of young stars also show high space velocities or peculiar motions
(Tetzlaff et al. 2010b), so-called run-away stars.
Two scenarios are accepted to produce run-away stars:  
(i) The binary-SN scenario (Blaauw 1961) is related to the formation of the high velocity 
NS: The run-away star and the NS are the products of a SN in a binary. 
The velocity of the former secondary is comparable to its original orbital velocity. 
(ii) The dynamical ejection scenario (Poveda et al. 1967) due to gravitational interactions between 
stars in dense clusters; an ejected massive star will later explode in a SN. 

There are $\sim 50$ young ($\le 15$ Myrs characteristic age), nearby ($\le 3$ kpc) 
NSs known, most of them radio pulsars,
plus a few radio-quiet thermal X-ray emitters (the M7),
for which both proper motion and distance are known (ATNF pulsar catalog
for radio pulsars and Table~1 for the M7).

By tracing back the 3D motion of all those NSs
and all known nearby young stellar associations,
we can find close encounters in space and time,
i.e. events, where NSs intersected with an association.
Then, the NS may have been born in that association at that time in a SN.

An additional encounter with a run-away star would provide additional evidence.
Hence, we have compiled a new catalog of young ($\le 50$ Myrs) run-away 
stars from Hipparcos (Tetzlaff et al. 2010b),
identified as such from their large peculiar spatial (3D), or transverse (2D), or any
1D velocity (e.g. radial velocity) or from their direction of motion compared
to neighbouring stars or associations.
In this way, first applied by us, we could identify $\sim 2500$ stars,
for which the probability for being a run-away star is larger than $50~\%$
(Tetzlaff et al. 2010b).

By tracing back the 3D space motion of not only all young nearby NSs
and all nearby young stellar associations,
but also also all known young run-away stars, 
we can find close encounters in space and time,
i.e. events, where a NS intersected with both an association
and a run-away star at the same place and the same time.
The time elapsed since the SN can then
be treated as (kinematic) age of that NS, which may well be a 
better age estimate for the NS than the characteristic age from
pulse rate and its derivative, which can be seen as upper limit.
The largest error source in calculating the flight path is the unknown 
radial velocity of the NSs, for which we use the most likely 
distribution derived from the 3D space velocity distribution 
(e.g. Arzoumanian et al. 2002, Hobbs et al. 2005) 
in a Monte-Carlo simulation.

The run-away O-star $\zeta$ Ophiuchi is an isolated 
main sequence star with a peculiar space velocity. Blaauw (1952)
suggested its origin in the Scorpius OB2 association due to 
its proper motion vector which points away from that association. 
Hoogerwerf et al. (2001; hereafter H01) investigated the 
origin of $\zeta$ Oph in more detail and proposed 
that it gained its high velocity in a binary SN in Upper Scorpius 
about $\sim 1$~Myr ago, which is also supported by its 
high helium abundance,
typical for run-away stars ejected by a SN in a binary.
They also identified a NS which might have been the former primary: 
PSR B1929+10 (PSR J1932+1059). We repeated the experiment of H01 
applying the same (and more recent) starting parameters for 
the pulsar with our software and could basically confirm the H01 results,
see Tetzlaff et al. (2010a). Hence, our precedure works well.

For the Guitar Nebula Pulsar, we could identify the 
Cygnus OB 3 association as birth place yielding a
radial velocity of $-27 \pm 75$ km/s (Tetzlaff et al. 2009), which 
is very close to the value found by the inclination of 
the bow shock around the Pulsar, the Guitar Nebula, namely near zero.

Among the M7, four objects have published distance estimates 
and proper motion values (Table~1) and thus their birth sites can in principle be found,
if assumptions on the radial velocity are made.
We give the details of the results in Tetzlaff et al. (2010a).
The SN distances are consistent with the Local Bubble and hence may have contributed to its 
formation or reheating (Bergh{\"o}fer \& Breitschwerdt 2002).
There is no doubt that the Sco OB2 
association to which Upper Scorpius belongs experienced some SNe in the past
(Bergh{\"o}fer \& Breitschwerdt 2002, Ma{\'{\i}}z-Apell{\'a}niz 2001, Fern{\'a}ndez et al. 2008).
There have been other investigations showing that many NSs may have been born 
within the Gould Belt (Perrot \& Grenier 2003).

Adopting a birth scenario of each NS, we can derive 
the kinematic NS age, which we 
always find to be lower than the characteristic spin-down age,
which is an upper limit to the true age. From cooling models we know that, 
given the surface temperature, for the NSs investigated here, the 
characteristic age is too high. This is also plausible since those objects 
are still very young. Thus, the lower kinematic ages better fit with cooling 
models (Tetzlaff et al. 2010a). 
The difference between the association age and the kinematic age of the NS
can be used to deduce the mass of the progenitor star, by the (well
justified) assumpion that all stars formed at roughly the same time:
The mass of the NS progenitor star is then the mass of a star with a life-time
being equal to the difference between the association age and the kinematic age of the NS.

\section{Finding the Neutron Star that was born in the SN that placed Fe60 on Earth}

For a recent (within a few Myrs) nearby SN (within 100 to 150 pc),
SN debris like Fe60 or Be10 should have been deposited 
in the Earth's crust (Ellis et al. 1996).
Such material has been detected (Raisbeck et al. 1987, Knie et al. 2004, Fitoussi et al. 2008).
The time since the SN is given by the flight time of the SN debris to Earth
and the time the Fe60 needs to be incorporated into the Earth crust, both quite uncertain.
Observationally, we can get a rough constrain also from the Fe60 life-time (Rugel et al. 2009).
If we can find the NSs formed by that SN, we can constrain its distance and age,
and maybe the mass (and, hence, debris yield) of the progenitor star,
important constraints for SN and nucleosynthesis models.

Next, we will trace back the flight path of all young nearby NSs,
all young nearby associations, and all run-away stars, in order to
find close encounters in space and time. We will also compare
the flight path of NSs with the Local Bubble and other nearby bubbles,
which may have been formed and/or reheated by SNe.
If we can find one or more NSs, which were formed
by a nearby (within roughly 100 pc) recent (within the last few Myrs) SN,
then this SN should have placed detectable debris like Fe60 on Earth.

The amount of, e.g., Fe60 debris found in the Earth crust is measured
and depends on the distance of the SN and the mass of the
progenitor star. Both are not known so far, so that models on SN
yields, which in some cases differ a lot (e.g. Timmes et al. 1995, 
Limongi \& Chieffi 2006, Frohlich et al. 2008, Fryer 2007),
could not be tested so far. 
If we can identify the NS born in such a SN, 
we know the distance towards the SN.
We may even be able to determine the mass of the progenitor star
from the difference between association age and NS flight time,
if all stars were formed at once in that association.
Such results would be very important and valuable for NS and SN theory
as well as for nucleosynthesis models, etc.

\end{document}